\documentclass{article}
\usepackage{graphicx} 
\usepackage{subfigure}
\usepackage{amsfonts}
\usepackage{amsmath}
\usepackage{multirow}
\usepackage{amsmath}
\usepackage{amssymb}
\usepackage{amsthm}
\usepackage{algorithm}
\usepackage{algorithmic}
\usepackage[round]{natbib}
\usepackage{color}

\usepackage{ijcai15}
\usepackage{graphicx}
\usepackage{times}

\title{CryptGraph: Privacy Preserving Graph Analytics on Encrypted Graph}
\author{Pengtao Xie and Eric Xing \\
School of Computer Science, Carnegie Mellon University\\
Pittsburgh, PA, 15213 \\
\{pengtaox,epxing\}@cs.cmu.edu}

\begin{document}
\newcommand{\mb}{\mathbf}
\maketitle

\begin{abstract}
  Many graph mining and analysis services have been deployed on the cloud, which can alleviate users from the burden of implementing and maintaining graph algorithms. However, putting graph analytics on the cloud can invade users' privacy. To solve this problem, we propose CryptGraph, which runs graph analytics on encrypted graph to preserve the privacy of both users' graph data and the analytic results. In CryptGraph, users encrypt their graphs before uploading them to the cloud. The cloud runs graph analysis on the encrypted graphs and obtains results which are also in encrypted form that the cloud cannot decipher. During the process of computing, the encrypted graphs are never decrypted on the cloud side.
  The encrypted results are sent back to users and users perform the decryption to obtain the plaintext results. In this process, users' graphs and the analytics results are both encrypted and the cloud knows neither of them. Thereby, users' privacy can be strongly protected. Meanwhile, with the help of homomorphic encryption, the results analyzed from the encrypted graphs are guaranteed to be correct. In this paper, we present how to encrypt a graph using homomorphic encryption and how to query the structure of an encrypted graph by computing polynomials. To solve the problem that certain operations are not executable on encrypted graphs, we propose hard computation outsourcing to seek help from users. Using two graph algorithms as examples, we show how to apply our methods to perform analytics on encrypted graphs. Experiments on two datasets demonstrate the correctness and feasibility of our methods.

\end{abstract}

\section{Introduction}
Recently, many cloud based graph analysis services have been launched, including IBM System G, Neo4j, GraphDB, Dydra, Infinity Graph, GraphLab, to name a few. In cloud based graph analytics, graph mining and analysis algorithms are deployed on cloud servers. To use the service, users upload their graphs to the cloud. The cloud runs graph algorithms over users' graphs and sends the results back to users. Putting graph analytics on the cloud has several benefits. First, users are alleviated from the burden of implementing and maintaining graph algorithms, which are time-consuming and error-prone. Second, many graph storage and management systems have been deployed on the cloud, such as cloud based graph database. Putting graph analytics also on the cloud makes it easier to integrate with other graph management services.
 
Despite the benefits provided by cloud graph analytics, it suffers a severe problem: invasion of users' privacy, which is also suffered by many other cloud based services. To use cloud graph analytics, users have to upload their graphs to the cloud, which completely exposes the graphs to malicious attackers and curious cloud owners. Besides, the results analyzed from users' input graphs are also visible to the cloud, which is again a huge threat of users' privacy. For some users, their graphs are so sensitive that their privacy cannot be compromised. Examples include atom graph of drugs \citep{dtpaids2004}, knowledge graph of confidential documents \citep{poh2012structured}, social graph containing sensitive information of individuals \citep{hay2008resisting,liu2008towards,akcora2012privacy}, to name a few. How to protect users' privacy is a vital factor and a big challenge for cloud graph analytics.

\begin{figure}
\center
\includegraphics[width=0.9\columnwidth]{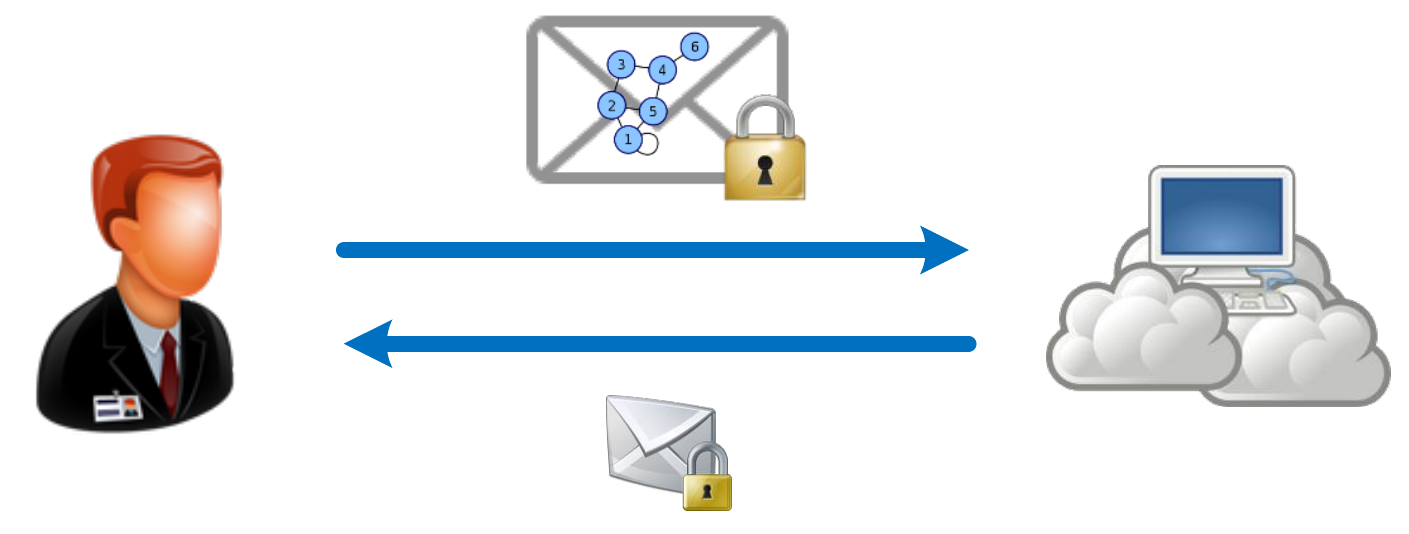}
\caption{Privacy-preserving graph analytics on encrypted graph. 
A user aims to analyze his graph through the cloud graph analytics. However, due to privacy concern, he refuses to let the cloud see the graph. He encrypts the graph before sending it to the cloud. The cloud runs graph analytics over the encrypted graph and obtains the results which are also ciphertexts that the cloud cannot decipher. The encrypted results are sent back to the user and the user does the decryption locally to obtain the readable plaintext results.  
}
\label{fig:cga}
\end{figure}

To solve this problem, we propose a solution called CryptGraph, which runs graph analysis algorithms on encrypted graphs. Users' graphs and the analysis results are both in encrypted form that the cloud cannot decipher, which ensures that users' privacy can never be leaked to the cloud. Figure \ref{fig:cga} illustrates our solution. Graph algorithms are deployed on the cloud. A user wants to analyze his graph through the cloud. Meanwhile, he refuses to let the cloud see the graph which leaks his privacy. He encrypts the graph and sends it to the cloud servers. The cloud receives the encrypted graph, runs graph analytics over it and obtains the results which are also in encrypted form that the cloud cannot decipher. During the process of computing, the encrypted graph is never deciphered on the cloud side. 
The ciphertext results are sent back to the user and the user does the decryption locally to obtain the readable plaintext results.
In this process, both the input graph and the output results are encrypted, thereby, the cloud has no chance to learn anything about the user. 
Users' privacy can be strongly guaranteed. Now the problem is, with the graph encrypted, how can we run analytics over it and ensure the results are correct? This turns out to be possible with the help of homomorphic encryption. Homomorphic encryption (HE) \citep{gentry2009fully} is an encryption scheme which allows certain computations over encrypted data. For instance, given that $x+y=z$ in the plaintext space, with HE, the equality still holds after encryption: $[\![ x ]\!] \bigoplus [\![ y ]\!]=[\![ z ]\!]$ , where $[\![ x ]\!]$, $[\![ y ]\!]$, $[\![ z ]\!]$ are the ciphertexts encrypting $x$, $y$, $z$ and $\bigoplus$ denotes homomorphic addition. Under homomorphic encryption, though data are encrypted, computations can still be performed and are guaranteed to be correct.
Practical homomorphic encryption schemes have certain restrictions over the computations they can perform. They can compute addition, subtraction, multiplication and negation, but do not support division and comparison. Many graph algorithms involve division and comparison, which makes it very challenging to perform analytics over encrypted graph. 

In this paper, we investigate how to encrypt a graph and how to perform computations over the encrypted graph. Our major contributions are:
\begin{itemize}
\item Our work is the first one proposing to encrypt a graph with homomorphic encryption, which can protect all the structure information of a graph without losing the ability to perform graph analytics over it.
\item We develop algorithmic and system solutions to adress the problem that division and comparison are not computable on encrypted graph. One is querying graph structure by computing polynomials, which avoids comparison. The other is hard computation outsourcing, which solicits users to perform a small fraction of operations which are hard for the cloud.
\item On two important graph algorithms and two datasets, we demonstrate that analytics on encrypted graph are correct and feasible.
\end{itemize}

%

The rest of the paper is organized as follows. Section 2 introduces related works and Section 3 introduces homomorphic encryption. Section 4 presents our method to run analytics on encrypted graphs and Section 5 studies two applications.
Section 6 presents experimental results and Section 7 concludes the paper.

\section{Related Works}
Preserving privacy of graph data has been extensively studied in many works. In general, they can be categorized into two paradigms: encryption based approaches and non-encryption based approaches. Encryption based approaches \citep{chase2010structured,cao2011privacy,poh2012structured,yi2014privacy} design encryption schemes to enable searching and querying over encrypted graphs. \citet{cao2011privacy} studied how to perform containment query, namely whether a graph is a subgraph of another graph. \citet{chase2010structured} developed a structural encryption scheme which supports neighbor queries and adjacency queries on encrypted graphs. \citet{poh2012structured} investigated how to perform searching over encrypted conceptual graphs. \citet{yi2014privacy} studied how to perform reachability queries: whether a vertex is reachable from another vertex. These encryption schemes are mainly designed to support graph searching and querying and they are unable to perform computation-oriented graph analytics such as computing clustering coefficient and PageRank. 
Several works \citep{duan2005privacy,sakuma2009link} utilized partially homomorphic encrytion to perform link analysis algorithms. Their problem settings are quite different from ours. In their settings, vertices of a graph belong to multiple parties and each party desires to protect its privacy against all others. In our setting, the graph is owned by one user who aims to protect the privacy of the graph against the cloud.

Non-encryption based approaches preserve graph privacy by anonymizing vertices and edges in the graph \citep{hay2008resisting,liu2008towards,zheleva2008preserving,zhou2008preserving} or using differentially private graph models \citep{sala2011sharing,task2012guide}. \citet{liu2008towards} proposed a $k$-degree vertex anonymization method under which for each vertex $v$ there exist at least $k-1$ other vertices in the graph with the same degree as $v$. \citet{hay2008resisting} anonymizes a graph by partitioning the vertices and then describing the graph at the level of partitions. \citet{sala2011sharing} developed a differentially private graph model for privacy protection, which generates a synthetic graph maintaining structural similarity to the original graph while introducing difference to provide the designed privacy guarantee. While these methods can protect the privacy of certain elements in the graph, such as vertices, they still expose a lot of information, such as structures, to the attackers. On the other hand, these methods change the original structure of the graph, thereby the analytics results obtained on the anonymized/synthesized graph might be quite different from those on the original graph.

\section{Homomorphic Encryption}
\label{sec:he}
\begin{figure}[t]
\center
\includegraphics[width=0.8\columnwidth]{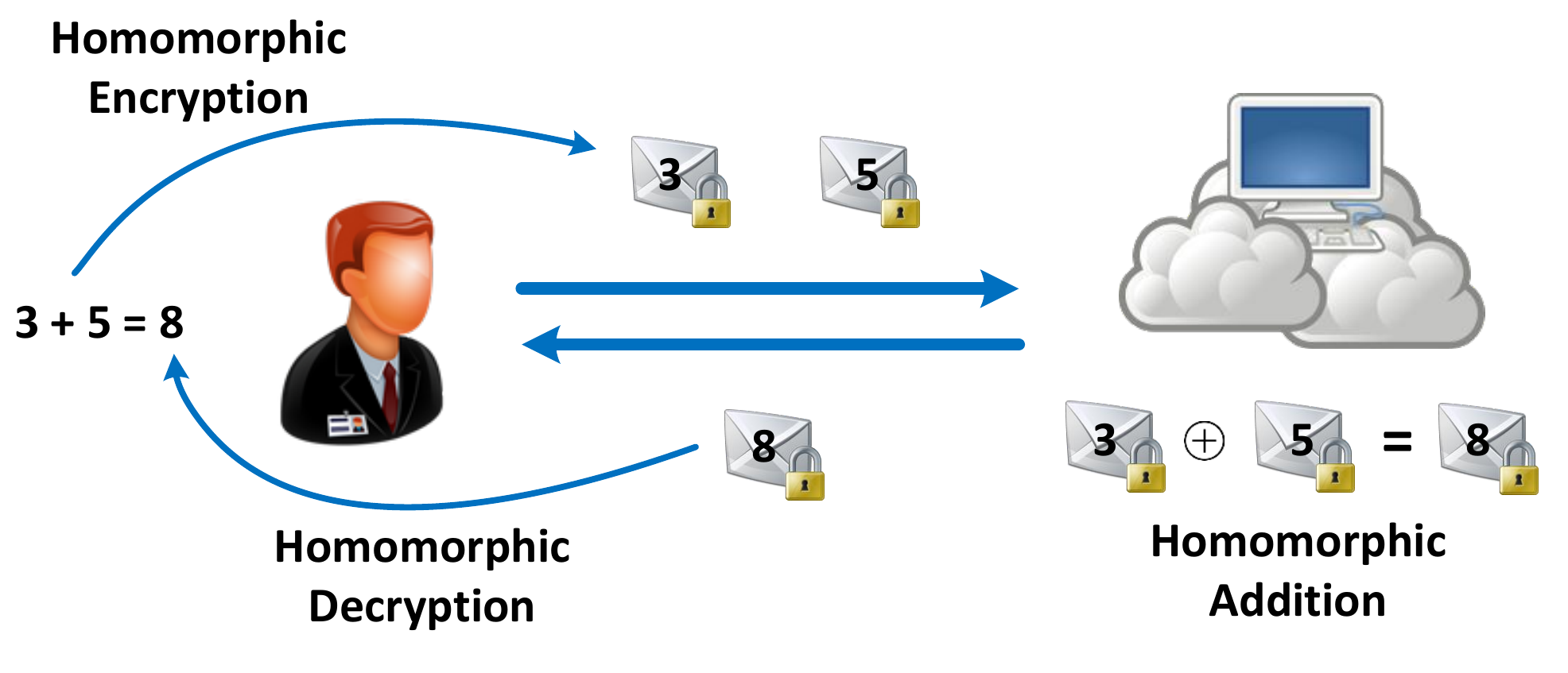}
\caption{An illustration of homomorphic encryption. 
The user wants to compute the sum of 3 and 5 without allowing the cloud to see these two numbers. He does the homomorphic encryption of 3 and 5, and sends the ciphertexts to the cloud. The cloud does a homomorphic addition of the two ciphertexts and gets a result which is also encrypted. The result is sent back to the user. The user does homomorphic decryption of the result and gets 8, which is exactly the sum of 3 and 5. 
}
\label{fig:he}
\end{figure}
In this section, we briefly introduce homomorphic encryption \citep{gentry2009fully} which allows certain computations over encrypted data. Figure \ref{fig:he} illustrates how homomorphic encryption works. The user wants to compute the sum of 3 and 5 using the cloud, but he is reluctant to let the cloud know these two numbers. He encrypts 3 and 5 into ciphertexts, sends them to the cloud and tells the cloud that he wants to add them. The cloud performs a $\oplus$ operation which is analogous to the addition on integers and obtains a result, which is also encrypted. The cloud sends the encrypted result back to the user. The user does homomorphic decryption locally and obtains the plaintext result 8, which is exactly the sum of 3 and 5. With homomorphic encryption, certain computations can successfully be performed although the data are in encrypted form. And the computation is guaranteed to be correct. 

In this paper, we use the ring based homomorphic encryption schemes \citep{brakerski2012fully,brakerski2012leveled,wu2012using,bos2013improved,lyubashevsky2013ideal}, which are arguably the most popular and practical ones. These schemes support addition, subtraction and multiplication on encrypted data, but cannot perform division and comparison. 
Besides, these schemes require the plaintext messages to be integers. If the message is a real number, we convert it into an integer by multiplying a large integer number $s$. $s$ is given to the cloud and is needed during homomorphic computation. After computation, the cloud sends the encrypted result together with a new scaling number $\hat{s}$ (which is related with $s$) back to the user. User does homomorphic decryption and divides the decrypted result (which is an integer) with $\hat{s}$ to obtain the real number result. 

\section{Graph Analytics on Encrypted Graph}
In this section, we present how to encrypt a graph and how to perform analytics on the encrypted graph. 

\subsection{Graph Representation and Encryption}
To enable management and computation of a graph on the cloud, it is necessary to let the cloud servers know certain information about the graph. Meanwhile, we protect the key information and prohibit the cloud servers from accessing it. In general, we expose the following information to the cloud: number of vertices in the graph, types of graph (undirected, directed, unweighted, weighted, bipartite). For bipartite graph, we also expose the number of vertices in each partite.
 The key information we want to keep private is: the meaning of vertices, degree of vertices, number of edges, whether an edge exists between two vertices, directions of edges, weights on edges. 

To protect the key information, we need to represent and encrypt the graph properly. 
We represent a graph with an adjacency matrix. 
Given a graph $G=(V, E)$ with vertices $V$ and edges $E$, we use an adjacency matrix $I$ of size $|V|\times|V|$ to represent its edges. $I_{ij}$ denotes the connectivity between vertex $i \in V$ and vertex $j \in V$. For simplicity, we assume there is at most one edge between each pair of vertices.
For an undirected graph, $I_{ij}=1$ if vertex $i$ and $j$ are connected; $I_{ij}=0$, otherwise.
 For a directed graph, 
$I_{ij}=1$ if the edge $E_{ij}$ is an inbound edge of vertex $i$; $I_{ij}=0$ if there is no edge between vertex $i$ and $j$; $I_{ij}=-1$ if $E_{ij}$ is an outbound edge of vertex $i$. In both the directed graphs and undirected graphs, $I_{ii}$ is set to 0.  
For a bipartite graph $G=(U,V, E)$, where $U$ and $V$ are the two disjoint vertex sets and $E$ denotes the edges, its adjacency matrix $I$ is of size $|U|\times |V|$. The rows of $I$ corresponds to vertices in $U$ and columns of $I$ corresponds to vertices in $V$. $I_{ij}=1$ if vertex $i\in U$ and $j\in V$ are connected; $I_{ij}=0$, otherwise. 
For a weighted graph, we use an additional matrix $W$ to represent weights on the edges, where $W_{ij}$ denotes the weight on the edge $E_{ij}$. Note that for a plaintext weighted graph, the weight matrix by itself can represent the graph structure where $W_{ij}=0$ denotes there is no edge between vertex $i$ and $j$ and the adjacency matrix is not needed. However, for an encrypted graph, the weight matrix cannot be used to depict graph structure  and the reason will be explained in Section 4.2.  

To encrypt a graph, we encrypt each element of the adjacency matrix $I$ and weight matrix $W$ independently using homomorphic encryption (HE). Let $\tilde{I}$ denote the encrypted adjacency matrix and $\widetilde{W}$ denote the encrypted weight matrix. HE is a random encryption scheme that adds random noise to plaintexts. Under random encryption, the same plaintext will be turned into different ciphertexts each time it is encrypted. Though the elements in plaintext $I$ can only take values in $\{0,1,-1\}$, the elements in ciphertext $\tilde{I}$ are different from each other (with high probability) due to the randomness of homomorphic encryption and the security guarantee that attackers cannot decipher these values is presented in \citep{brakerski2012fully,brakerski2012leveled,wu2012using,bos2013improved,lyubashevsky2013ideal}. In CryptGraph, all the cloud can see are $\tilde{I}$ and $\widetilde{W}$, where the elements are ciphertexts which make no sense to the cloud. The cloud has no way to know whether an edge exists between two vertices or the direction and weight of an edge, thereby it knows nothing about the structure of the graph. 

\subsection{Query Graph Structure by Computing Polynomials }
Querying the structure of a graph is a major operation in graph algorithms, such as selecting the neighboring vertices of a given vertex, selecting the inbound or outbound edges of a vertex, judging whether a group of vertices form a strong clique or not, etc. On unencrypted graphs, this can be easily done. To represent neighborhood relationships between two vertices or between an edge and a vertex, we can use an adjacency list. To judge whether a group of vertices are interconnected, we can use \textit{if-else} statement to inspect the connectivities among these vertices. However, on encrypted graphs, this becomes very hard. First, an adjacency list cannot be used since it leaks the structure information of a graph. Second, \textit{if-else} judgment cannot be performed since comparison is not executable over encrypted data. 

To solve this problem, we propose to rewrite all queries into polynomial computations. Polynomials involve addition and multiplication, which are computable by the homomorphic encryption scheme. We assume all queries are boolean, where the input is a statement and the output is either \textit{true} (denoted with 1) or \textit{false} (denoted with 0). According to how many vertices are involved, we categorize queries into two kinds: atom query which is about the relationship between two vertices and conjunction query which is about the relationship between multiple vertices. 
Atom queries are of three kinds: 1, are vertex $i$ and vertex $j$ connected with an undirected edge? 2, is vertex $j$ the inbound neighbor of vertex $i$? (equivalently, is there a directed edge from $j$ to $i$?) 3, is vertex $j$ the outbound neighbor of vertex $i$? (equivalently, is there a directed edge from $i$ to $j$). Conjunction queries can be constructed by conjoining atom queries. For instance, ``are vertex $i$,$j$,$k$ in a triangle?'' can be equivalently written as ``are $i$ and $j$ connected? \textit{AND} are $j$ and $k$ connected? \textit{AND} are $i$ and $j$ connected?''. The answer of a conjunction query is \textit{true} if and only if the answers of all atom queries are \textit{true}.
On encrypted graphs, we construct polynomials $p:\tilde{S}\to\{\tilde{1},\tilde{0}\}$ to answer these queries, where $\tilde{S}$ is a subset of elements in the adjacency matrix $\tilde{I}$. 
Note that we use the notation $\tilde{1},\tilde{0}$ to emphasize that the outputs of polynomials are also encrypted and the cloud has no way to decipher them. In the rest of the paper, if not explicitly stated, $\tilde{a}$ denotes the encrypted ciphertext of $a$.
For atom queries involving vertex $i$ and $j$, $\tilde{S}=\{\tilde{I}_{ij}\}$.
For conjunction queries involving a set of vertices $A$, $\tilde{S}=\{\tilde{I}_{ij}|i\in A \wedge j\in A \wedge i\neq j\}$. For the first atom query ``are vertex $i$ and vertex $j$ connected with an undirected edge?'', it can be directly answered from the adjacency matrix $\tilde{I}$ of an undirected graph. $\tilde{I}_{ij}=\tilde{1}$ means there is an edge between $i$ and $j$;  $\tilde{I}_{ij}=\tilde{0}$, otherwise. So the polynomial would simply be $p_{1}(x)=x$. The other two queries can be answered by evaluating polynomials over the adjacency matrix $\tilde{I}$ of directed graphs. To answer ``is vertex $j$ the inbound neighbor of vertex $i$?'', we evaluate $p_{2}(x)=\frac{1}{2}x^{2}+\frac{1}{2}x$ over $\tilde{I}_{ij}$. Recall that $\tilde{I}_{ij}=\tilde{1}$ denotes there is a directed edge from $j$ to $i$ and $\tilde{I}_{ij}=\tilde{0}$ otherwise. $p_{2}(x)$ maps $\tilde{1}$ to $\tilde{1}$ and maps $\tilde{0}$ and $\widetilde{-1}$ to $\tilde{0}$. Thereby, $p_{2}(x)$ is capable to pick out the inbound edges. Similarly, to answer ``is vertex $j$ the outbound neighbor of vertex $i$?'', we evaluate $p_{3}(x)=\frac{1}{2}x^{2}-\frac{1}{2}x$ over $\tilde{I}_{ij}$.
This polynomial is compatible with the representation of outbound edges. We call these polynomials answering atom queries as atom polynomials. Note that $p_{2}(x)$ and $p_{3}(x)$ involve real numbers $\frac{1}{2}$ and $-\frac{1}{2}$, which are not directly computable under homomorphic encryption. We use the method discussed in Section \ref{sec:he} to deal with these real numbers. In the rest of the paper, we tackle all real numbers in this way without explicitly reiterating. In Section 4.1, we state that the weight matrix cannot be used to represent the structure of a weighted graph. This is because we cannot find atom polynomials to map the real-valued weights to $\{1,0\}$.

Knowing how to answer atom queries, we can answer conjunction queries by multiplying atom polynomials. A conjunction query consisting $K$ atom queries $Q^{(c)}=Q^{(a)}_{1}\wedge Q^{(a)}_{2}\wedge \cdots \wedge Q^{(a)}_{K}$ can be answered with a polynomial $p^{(c)}=p^{(a)}_{1}\cdot p^{(a)}_{2}\cdots \cdot p^{(a)}_{K}$, which is the product of $K$ atom polynomials. $p^{(c)}$ is called conjunction polynomial.
$p^{(c)}$ equals to $\tilde{1}$ if and only if all atom polynomials equal to $\tilde{1}$, this is consistent with the fact that the answer of a conjunction query is \textit{true} if and only if the answers of all atom queries are \textit{true}.

 Graph algorithms usually involve operations over a set where elements satisfy certain conditions. For example, to count the triangles containing vertex $i$, we need to calculate the cardinality of a set $\{(i,j,k)|\textrm{vertexe}\mspace{3mu} i,j,k \mspace{3mu} \textrm{form a triangle}\}$. In PageRank, to update the PageRank value for a vertex $i$, we need to sum up all the weighted PageRank values of the inbound neighbor set $\{j| j \mspace{3mu}\textrm{is the inbound neighbor of}\mspace{3mu} i\}$. The cloud can answer the queries ``do vertices $i$,$j$,$k$ form a triangle?'' and ``is $j$ the inbound neighbor of $i$?'' by calculating polynomials. However, the cloud does not know what the answers are since they are also encrypted. Thereby, the cloud has no way to create a set whose elements meet certain conditions. To solve this problem, we propose to perform masked computations over elements in the universal set. Operation on each element is masked with the authenticity that whether the element satisfies a certain condition. If the authenticity is $\tilde{1}$, the element contributes to the final result. If the authenticity is $\tilde{0}$, the element has no influence of the final result. For example, in PageRank, to update the PageRank value of a vertex $i$, one needs to sum up the weighted PageRank values $w_{j}$ of its inbound neighbors $\mathcal{N}(i)$. Since the cloud cannot construct $\mathcal{N}(i)$, we can perform the masked summation $\sum_{j\in V} \tilde{w}_{j}\cdot p_{2}(\tilde{I}_{ij})$ over all vertices $V$, where $p_{2}(x)$ is the atomic polynomial selecting inbound neighbors. If $j$ is an inbound neighbor of $i$, $p_{2}(\tilde{I}_{ij})$ equals to $\tilde{1}$ and the contribution $\tilde{w}_{j}$ of vertex $j$ will be counted into the final sum. If $j$ is not an inbound neighbor of $i$, $p_{2}(\tilde{I}_{ij})=\tilde{0}$ and $\tilde{w}_{j}$ will not be contributed to the final sum.


\subsection{Hard Computation Outsourcing}
\begin{figure}
\center
\includegraphics[width=0.9\columnwidth]{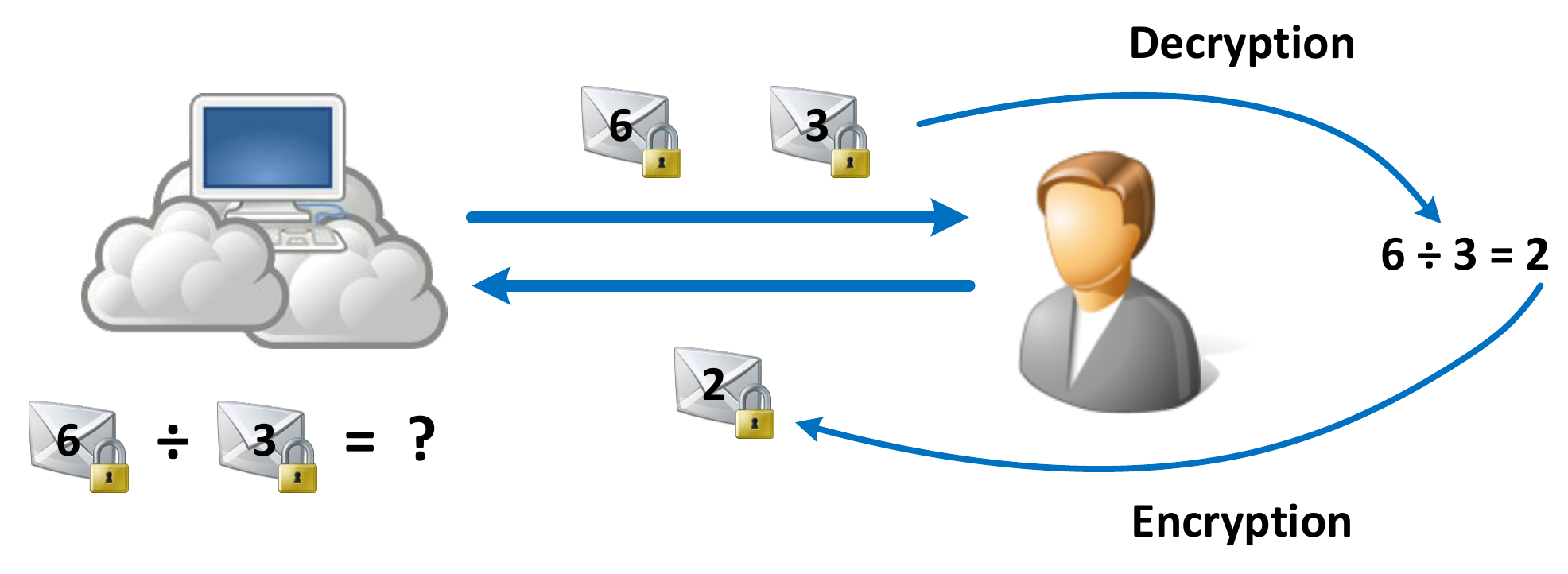}
\caption{An illustration of hard computation outsourcing. 
The cloud is not able to compute the division between encrypted 6 and 3. It sends the two ciphertexts back to the user and asks the user to do a division. The user decrypts the ciphertexts and gets the plaintext numbers 6 and 3. User divides 6 by 3 and gets 2, which is very easy to compute. The result 2 is encrypted and sent back to the cloud. The cloud gets the encrypted 2 and resumes the computation.
}
\label{fig:hard_outsource}
\end{figure}

Recall that homomorphic encryption cannot support division and comparison, which we refer to as hard computations. These two types of computations exist extensively in graph algorithm. For example, comparison is needed in single source shortest paths (SSSP) and division is required in PageRank. While these two operations are hard for homomorphic encryption schemes, they are easy for users. Thereby, we propose to outsource the hard computations back to the users who own the graphs. Whenever the cloud encounters a hard computation $O$ over some data $\tilde{D}$, it sends $\tilde{D}$ back to the graph owner and solicits the owner to perform $O$ on $\tilde{D}$. Note that, $\tilde{D}$ are intermediate results which are in encrypted form. On users' side, there is a client process in charge of processing the hard computations for the cloud. The client first decrypts $\tilde{D}$ into plaintext data $D$, then performs $O$ over $D$ and obtains a result $R$. $R$ is then encrypted into $\tilde{R}$ which is sent back to the cloud. The cloud receives $\tilde{R}$ and resumes the computation. Figure \ref{fig:hard_outsource} illustrates the idea of hard computation outsourcing. The cloud wants to compute the division between two ciphertexts. Unfortunately, division is not computable over encrypted data. The cloud sends the two ciphertexts back to the user and asks the user to perform a division. The user decrypts the ciphertexts and obtains two numbers 6 and 3. The user divides 6 by 3 and obtains 2, which is very easy to compute. The result 2 is encrypted and sent back to the cloud. The cloud receives the encrypted 2 and resumes the computation. 

The drawback of hard computation outsourcing is that users are involved in the computation. This in some sense violates the motivation of cloud graph analytics, which is to alleviate users' burden of implementing and maintaining their own graph algorithms. To mitigate this drawback, we propose two principles. First, hard computations should be a small proportion of the whole computations in a graph algorithm. In other words, most of the computations should happen on the cloud side and users only need to perform a few computations that the cloud cannot do. 
Second, the computations that users help to perform should not be algorithm-specific. 
Users only need to provide a set of basic computations which are common to all graph algorithms. This ensures that the computations on users' side are stable no matter how significantly the graph algorithms change on the cloud side.

\section{Applications}
In this section, we show how to perform graph analytics on encrypted graphs with the above mentioned two techniques. Specifically, we study two graph algorithms: computing clustering coefficient and PageRank.

\subsection{Clustering Coefficient}
Clustering coefficient \citep{suri2011counting} $cc(i)$ of a vertex $i$ in an undirected graph measures how tightly-knit the community is around the vertex. It is defined as
\begin{equation}
cc(i)=\frac{|(j,k)\in E|j\in \mathcal{N}(i) \mspace{3mu}\wedge \mspace{3mu} k \in \mathcal{N}(i)|}{0.5\cdot d_{i}\cdot (d_{i}-1)}
\end{equation}
where $\mathcal{N}(i)$ is the neighbor set of vertex $i$ and $d_{i}$ is the degree of vertex $i$. The numerator is basically the count of triangles containing vertex $i$. To compute the clustering coefficients of vertices on an encrypted graph, the cloud first computes the degree $\tilde{d}_{i}$ of each vertex, which is the sum of all elements of the $i$th row in $\tilde{I}$. 
Then it computes the denominator $\widetilde{0.5}\cdot \tilde{d}_{i}\cdot (\tilde{d}_{i}-\tilde{1})$. Since division is not supported, the cloud can use hard computation outsourcing to compute the reciprocal of $\widetilde{0.5}\cdot \tilde{d}_{i}\cdot (\tilde{d}_{i}-\tilde{1})$. 
In this case, users only need to compute $N$ divisions, where $N$ is the number of vertices. This amount of computation is light-weight and meets the first principle of hard computation outsourcing. 
Using the polynomial based structure querying method, the triangle count on the numerator can be equivalently written into
\begin{equation}
\begin{array}{l}
|(j,k)\in E|j\in \mathcal{N}(i) \mspace{3mu}\wedge \mspace{3mu} k \in \mathcal{N}(i)|\\
=\sum\limits_{\substack{
   j=1 \\
   j\ne i
  }}^{|V|}
  \sum\limits_{\substack{
   k=j+1 \\
   k\ne i
  }}^{|V|}
  \tilde{I}_{ij}\cdot \tilde{I}_{ik}\cdot \tilde{I}_{jk}
\end{array}
\end{equation}
where the equality holds subject to encryption and decryption. The conjunction query $(j,k)\in E \wedge j\in \mathcal{N}(i)  \wedge k \in \mathcal{N}(i)$ is evaluated with a conjunction polynomial $p(x)=x\cdot x\cdot x$.
This equation basically examines all possible $(j,k)$ pairs to see whether they form a triangle with $i$. $(i,j,k)$ forms a triangle if and only if $\tilde{I}_{ij}=\tilde{1}$, $\tilde{I}_{ik}=\tilde{1}$ and $\tilde{I}_{jk}=\tilde{1}$, equivalently, $\tilde{I}_{ij}\cdot \tilde{I}_{ik}\cdot \tilde{I}_{jk}=\tilde{1}$.

\subsection{PageRank}
PageRank \citep{page1999pagerank} is an algorithm to compute the importance of vertices in a directed and unweighted graph. The underlying assumption is that more important vertices are likely to receive more links (inbound edges) from other vertices. The PageRank value $PR(i)$ of vertex $i$ can be computed as
\begin{equation}
\label{eq:pr}
PR(i)=\frac{1-d}{N}+d\sum\limits_{j\in NI(i)}\frac{1}{DO(j)}PR(j)
\end{equation}
where $NI(i)$ are the inbound neighbors of $i$ and $DO(j)$ is the outdegree of $j$. $d$ is a parameter which we set to 0.85 and $N$ is the total number of vertices. This equation is iteratively computed over each vertex until the PageRank value of each vertex converges.

To run PageRank over an encrypted graph, the cloud needs to compute Eq.(\ref{eq:pr}) under the homomorphic encryption scheme. In this equation, $(1-d)/N$ and $1/DO(j)$ involve divisions, which are not supported by the encryption scheme. Since the cloud knows the fixed parameter $d$ and the number of vertices $N$, it can compute $(1-d)/N$ on plaintext data. To compute $1/\widetilde{DO}(j)$, the cloud uses hard computation outsourcing. The cloud first computes $\widetilde{DO}(j)$ for each vertex, then sends $\widetilde{DO}(j)$ to the graph owner and solicits a \textit{reciprocal} operation. Users decrypt $\widetilde{DO}(j)$ into plaintext value $DO(j)$, compute the reciprocal $1/DO(p_{j})$, encrypt the reciprocal and send the ciphertext back to the cloud.
In the whole process, users only need to compute the reciprocal of $N$ numbers once, which is light-weight. Computing $\sum_{j\in NI(i)}\frac{1}{DO(j)}PR(j)$ requires to pick out the inbound neighbors of $i$. To achieve this, the cloud evaluates the atom polynomial $p(x)=\frac{1}{2}x^{2}+\frac{1}{2}x$ on each element $\tilde{I}_{ij}$ of the $i$th row of the adjacency matrix $\tilde{I}$. If $p(\tilde{I}_{ij})=\tilde{1}$, $j$ is an inbound neighbor of $i$. Since the cloud cannot know the output of $p(\tilde{I}_{ij})$, it performs the masked summation over all vertices
\begin{equation}
\label{eq:pr_rewrite}
\sum\limits_{j\in NI(i)}\frac{1}{DO(j)}PR(j)=\sum\limits_{j=1}^{N}p(\tilde{I}_{ij}) \frac{1}{\widetilde{DO}(j)}\widetilde{PR}(j)
\end{equation}
In this equation, only inbound neighbors where $p(\tilde{I}_{ij})=\tilde{1}$ contributes to the final sum.

\begin{table}[t]
\caption{Clustering coefficients (CC) of Dolphin Social Network}
\label{table:cc_dolphin}
\begin{center}
\begin{tabular}{c|c|c|c|c}
\hline
Graph& Avg CC& Max CC& Min CC& MSE\\
\hline
Unencrypted&0.2590&0.6667&0& 0\\
Encrypted&0.2589&0.6666&0& 1.7e-9\\
\hline
\end{tabular}
\end{center}
\end{table} 
\begin{table}[t]
\caption{PageRank (PR) values of Political Blogs}
\label{table:pr_political}
\begin{center}
\begin{tabular}{c|c|c|c|c}
\hline
Graph& Avg PR& Max PR& Min PR&MSE\\
\hline
unencrypted&0.00044&0.0878&0.0001& 0\\
encrypted&0.00043&0.0875&0.0001&1.0e-7 \\
\hline
\end{tabular}
\end{center}
\end{table}

\begin{table}[t]
\caption{Time of triangle counting for each vertex in Dolphin Social Network}
\label{table:time_dolphin}
\begin{center}
\begin{tabular}{c|c}
\hline
Graph& Time (s)\\
\hline
unencrypted&1e-8\\
encrypted& 18.77\\
\hline
\end{tabular}
\end{center}
\end{table}

\begin{table}[ht]
\caption{Time of updating PageRank value for each vertex in Political Blogs}
\label{table:time_political}
\begin{center}
\begin{tabular}{c|c}
\hline
Graph& Time (s)\\
\hline
unencrypted&2e-7\\
encrypted& 56.43\\
\hline
\end{tabular}
\end{center}
\end{table} 
\section{Experiments}
In this section, we evaluate the correctness and speed of graph analytics on two encrypted graphs.

\subsection{Experimental Setup}
We use two datasets in the experiments. The first one is the Dolphin Social Network \citep{lusseau2003bottlenose} which is an undirected social network of frequent associations between 62 dolphins. The graph has 62 nodes and 159 edges. The second dataset is the Political Blogs \citep{adamic2005political} which is a directed network of hyperlinks between weblogs on US politics.
This graph has 1490 nodes and 16718 edges. Some edges are removed to ensure there is at most one edge between each pair of nodes.
We used the homomorphic encryption (HE) scheme implemented\footnote{https://github.com/dwu4/fhe-si} by \citep{wu2012using} and follow their parameter setting of the encryption scheme.
\subsection{Result}
On each of the datasets, we perform graph analytics on the plaintext graph and the encrypted graph and compare their difference. The graph analytics we performed are: 1, compute the clustering coefficient of each vertex in the Dolphin Social Network; 2, compute the PageRank value of each vertex in the Political Blogs graph.
Table \ref{table:cc_dolphin} shows the average, max and min clustering coefficients on the plaintext and encrypted Dolphin Social Network. We also show the mean square error (MSE) $\frac{1}{N}\sum\limits_{n=1}^{N}||CC^{(p)}_{n}-CC^{(e)}_{n}||^{2}$ between the clustering coefficients $\{CC_{n}^{(p)}\}_{n=1}^{N}$ computed on the plaintext graph and $\{CC_{n}^{(e)}\}_{n=1}^{N}$ obtained from the encrypted graph, where $N$ is the number of vertices. From Table \ref{table:cc_dolphin}, we can see that the results obtained from the encrypted graph are very close to those obtained from the plaintext graph. Through the graph is encrypted, clustering coefficients can still be computed and are guaranteed to be correct. Encrypting the graph does not incur significant performance loss. The difference between the clustering coefficients on the plaintext graph and the encrypted graph mainly comes from the conversion of real numbers to integers when doing the computations on encrypted data.
Table \ref{table:pr_political} shows the average, max and min PageRank values on the unencrypted and encrypted Political Blogs graph. We also compute the MSE $\frac{1}{N}\sum\limits_{n=1}^{N}||PR^{(p)}_{n}-PR^{(e)}_{n}||^{2}$ between $\{PR_{n}^{(p)}\}_{n=1}^{N}$ and $\{PR_{n}^{(e)}\}_{n=1}^{N}$ which are the PageRank values computed from the plaintext graph and the encrypted graph respectively and $N$ is the number of vertices. From Table \ref{table:pr_political}, we can see that the PageRank values obtained from the encrypted graph are very close to those obtained from the plaintext graph. Again, encrypting the graph does not degrade the performance of PageRank. Table \ref{table:cc_dolphin} and Table \ref{table:pr_political} indicate that graph analytics on encrypted graphs are guaranteed to be correct.

To check the speed of graph analytics on the encrypted graph, we measure the time of the key operations in each analytic task. In computing clustering coefficients, the key operation is triangle counting. Table \ref{table:time_dolphin} shows the average time (in seconds) of triangle counting for each vertex in the Dolphin Social Network. In PageRank, updating PageRank value for each vertex using Eq.(\ref{eq:pr}) and Eq.(\ref{eq:pr_rewrite}) is the key operation. Table \ref{table:time_political} shows the average time (in seconds) of updating PageRank value for each vertex in Political Blogs. Time spent on other operations such as encryption, decryption, outsourcing (communication between user clients and cloud servers) is neglectable compared with homomorphic computatons, thereby, we omit it here.
As can be seen from Table \ref{table:time_dolphin} and \ref{table:time_political}, not surprisingly, the computations on encrypted graph are much slower than those on the plaintext graph. However, considering that most graph analytics do not need to be done in realtime, the time spent on encrypted graph is acceptable. Moreover, computations on encrypted graphs can be easily parallelized on the multi-core and distributed computing facilities on cloud platforms, which can improve the speed in great manner and make encrypted graph analytics more feasible.

\section{Conclusion}
In this paper, we study how to perform graph analytics on encrypted graphs, with the aim to protect users' privacy. We investigate how to encrypt a graph such that its structural information is protected from leakage. Given a encrypted graph, we propose to query its structure by computing polynomials. Considering that division and comparison are not computable by homomorphic encryption, we propose hard computation outsourcing, which solicits graph owners to help compute a small fraction of operations that the cloud cannot do. Using two graph algorithms as examples, we show how to leverage polynomial based structure querying method and hard computation outsourcing to conduct graph analysis on encrypted graphs. Experiments on two real datasets demonstrate that analytics on encrypted graph are correct and feasible.

\bibliographystyle{named}
\bibliography{refs}

\end{document}